\documentclass[prb,aps,twocolumn,showpacs,superscriptaddress]{revtex4}
\usepackage{amsmath,amssymb}
\usepackage{graphicx}
\begin{document}

\title{Evidence of a small crystal field anisotropy in GdCoIn$_5$}
\author{D. Betancourth}
\affiliation{Centro At\'omico Bariloche (CNEA) and Instituto Balseiro (U. N. Cuyo), 8400 Bariloche, R\'io Negro, Argentina}

\author{D. J. Garc\'ia}
\affiliation{Centro At\'omico Bariloche (CNEA) and Instituto Balseiro (U. N. Cuyo), 8400 Bariloche, R\'io Negro, Argentina}

\author{V. F. Correa}
\affiliation{Centro At\'omico Bariloche (CNEA) and Instituto Balseiro (U. N. Cuyo), 8400 Bariloche, R\'io Negro, Argentina}

%\author{Orazi, Paredes, Sposetti, Calder\'on}
\date{\today}



\begin{abstract}

We investigate the effects of an applied magnetic field on the magnetic properties of the antiferromagnet GdCoIn$_5$. The prominent anisotropy observed in the susceptibility below $T_N$ is rapidly suppressed by a field of just a few Tesla. Further evidence of this low energy-scale is obtained from magnetoresistance and magnetostriction experiments. The lattice lenght, particulary, shows a sudden change below 2 Tesla when the magnetic field is applied perpendicular to the crystallographic $\hat{c}$-axis. The experimental results as a whole can be attributed to a small but non negligible higher-order crystalline electric field. 

\end{abstract}

\maketitle

\section{Introduction}

GdCoIn$_5$ is member of the extensively studied 115 family of compounds RTIn$_5$ (R = rare earth, T = Co, Rh, Ir). Different experiments\cite{Betancourth2014} show a second-order phase transition to an antiferromagnetic state at $T_N$ = 30 K. 
Magnetic ground states are also observed in several other members of the 115 family\cite{Isikawa2004,Inada2006,Hudis2006,Huy2009}. In general, these magnetic states are influentiated by the crystal electric field (CEF) produced by the surrounding ions\cite{vanHieu2007}. In this sense, one expects a different situation for GdCoIn$_5$ since the half-filled 4$f$ orbital of the Gd$^{3+}$ ion has zero orbital momentum wich makes the CEF effects much less important. This would imply, for instance, that anisotropy in the magnetic properties should be negligible. Nonetheless, very recently\cite{Betancourth2014} we have shown that an important anisotropy is observed in the magnetic susceptibility below $T_N$.
The susceptibility indeed shows an easy magnetic axis along the basal ab-plane of the tetragonal crystal structure.

Here we show that the magnetic anisotropy of GdCoIn$_5$ has a small characteristic energy-scale: the magnetic susceptibility becomes mostly isotropic under an applied field $B$ of a few Tesla. This is further confirmed by some singularities observed in the magnetoresistance and magnetostriction around 2 Tesla when $B \perp$ [001]. The experimental evidence points towards a small but non-negligible higher-order CEF effect as the origin of this low energy-scale.
  
\section{Experimental details}

Single crystalline samples of GdCoIn$_5$ were grown by the self-flux technique as described elsewhere\cite{Betancourth2014}. The magnetization $M$ was measured in a Quantum Design MPMS magnetometer. A high-resolution capacitive dilatometer\cite{Schmiedeshoff2006} was used in the magnetostriction experiments, while a standard four probe setup was used in the magnetoresistance experiments.   

\section{Results}

\begin{figure}[h]
\begin{center}
\includegraphics[width=\columnwidth]{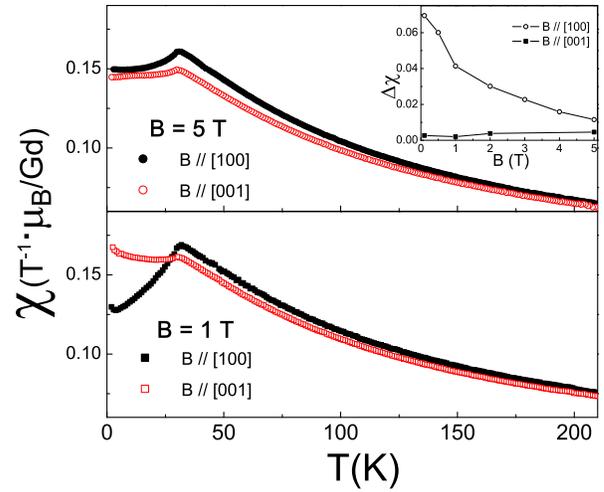}
\end{center}
\caption{(Color online) Temperature dependence of the magnetic susceptibility along the [100] and [001] directions in an applied field $B =$ 1 T (lower panel) and $B =$ 5 T (upper panel). Inset: susceptibility difference between its value at $T_N$ and its minimum value below $T_N$.} 
\label{fig1}
\end{figure}

Figure \ref{fig1} shows the temperature dependence of the magnetic static susceptibility ($\chi = M/B$) parallel and perpendicular to the [001] direction in an applied magnetic field $B =$ 1 T (lower panel) and $B =$ 5 T (upper panel). The transition to the antiferromagnetic state is detected as a peak in both directions ($T_N \approx$ 30 K). At $B =$ 1 T, the anisotropy observed below $T_N$ is typical of an antiferromagnet with ordered moments lying perpendicular to [001]. Remarkably, however, this pronounced anisotropy is significantly reduced at $B =$ 5 T. 
A quantitative estimate of this field-induced ``isotropization'' is shown in the inset of Fig. \ref{fig1}. It depicts the susceptibility difference between its value at $T_N$ and its minimum value below $T_N$, in both directions. While $\Delta \chi_{001}$ is very small and field independent, $\Delta \chi_{100}$ is quite large at low fields but it rapidly decreases above $B \sim$ 2 T.

\begin{figure}[]
\begin{center}
\includegraphics[width=\columnwidth]{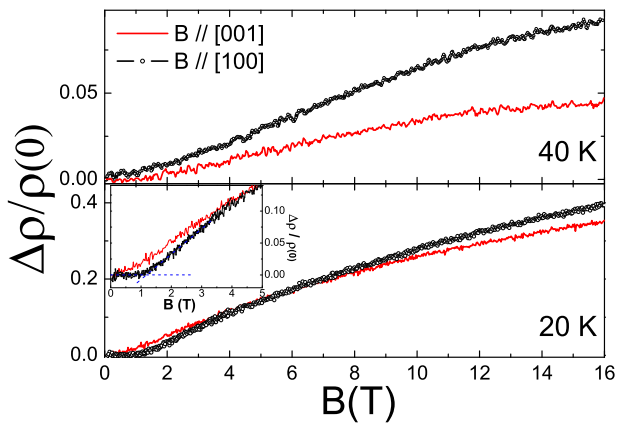}
\end{center}
\caption{(Color online) Magnetoresistance as function of a magnetic field applied along the [100] and [001] directions at $T$ = 20 K (lower panel) and $T$ = 40 K (upper panel). Inset: Low field magnetoresistance at 20 K. Dashed lines are guides to the eye.}
\label{fig2}
\end{figure}

Another evidence of this low energy-scale comes from the ab-plane magnetoresistance $\Delta \rho\left(B\right)$/$\rho\left(0\right)$. Figure \ref{fig2} displays $\Delta \rho\left(B\right)$/$\rho\left(0\right)$ at two different temperatures, above ($T$ = 40 K) and below ($T$ = 20 K) the ordering temperature and for $B$ along the [100] and [001] directions. The magnetoresistance is positive as expected for predominant antiferromagnetic correlations and it is progressively reduced above $T_N$. The inset of Fig. \ref{fig2} shows a detailed view of the low field magnetoresistance at 20 K. It can be seen that when $B \parallel$ [100], $\Delta \rho\left(B\right)$/$\rho\left(0\right)$ is zero for $B \lesssim$ 1 T. The effect disappears above $T_N$.

\begin{figure}[]
\begin{center}
\includegraphics[width=0.90\linewidth,keepaspectratio]{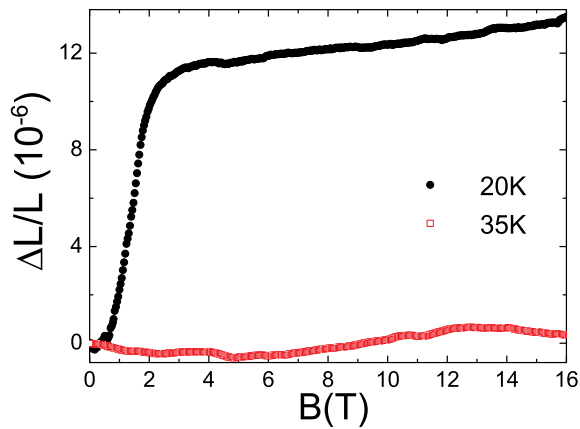}
\caption{(Color online) Longitudinal magnetostriction along [100] at $T$ = 20 K and $T$ = 35 K.}
\end{center}
\label{fig3}
\end{figure}

But the most notable evidence of a low energy-scale is obtained from magnetostriction. Figure \ref{fig3} shows the field dependence of the longitudinal magnetostriction along the [100] direction at two different temperatures. Below $T_N$, the lattice length shows an abrupt increase around 2 T (see the curve at 20 K). This effect becomes less important as the temperature is raised and it finally disappears at $T_N$. Above $T_N$ the magnetostriction is much smaller and shows a smooth field dependence (see the curve at 35 K). On the other hand, no peculiarities in the striction is observed when the magnetic field points in the [001] direction.    

Such a small energy-scale is suggestive of a tiny but non-negligible higher order crystalline electric field acting on the Gd$^{3+}$ ions producing slightly anisotropic interactions. This CEF is responsible of the orientation of the magnetic moments along the basal plane in the ordered state. 

The magnetic energy associated with an applied magnetic field $B \approx$ 2 T is enough to overcome the CEF, hence suppressing the magnetic anisotropy and, eventually, inducing a change in the relative orientation of neighboring magnetic moments (i.e., a change in the magnetic correlations) which causes the observed magnetostriction\cite{Callen1965}.

\section{Conclusions}

Even though the antiferromagnetic state of GdCoIn$_5$ is very robust against an applied magnetic field\cite{Betancourth2014}, different experiments like magnetic suscpetibility, resistivity and magnetostriction show evidence of a low energy-scale below $T_N$. Such a low energy-scale is compatible with a small crystalline electric field (CEF) acting on the Gd$^{3+}$ ions. This CEF is responsible of the observed low temperature anisotropy in the magnetic susceptibility.

\begin{acknowledgements}
We thank Pablo Cornaglia for fruitful discussions. Work partially supported by CONICET and SeCTyP-UnCuyo from Argentina.

\end{acknowledgements}


\begin{thebibliography}{}

\bibitem{Betancourth2014}  D. Betancourth, J. I. Facio, P. Pedrazzini, C. B. R. Jesus, P. G. Pagliuso, V. Vildosola, P. S. Cornaglia, D. J. Garc\'ia, and V. F. Correa, \textit{arXiv:1406.1536v2 [cond-mat.str-el]}.

\bibitem{Isikawa2004}  Y. Isikawa, D. Kato, A. Mitsuda, T. Mizushima, and T. Kuwai, \textit{J. Magn. Magn. Mater.} \textbf{272-276}, 635 (2004).

\bibitem{Inada2006}  Y. Inada, M. Hedo, T. Fujiwara, T. Sadamasa, and Y. Uwatoko, \textit{Physica B} \textbf{378-380}, 421 (2006). 

\bibitem{Hudis2006}  J. Hudis, R. Hu, C. L. Broholm, V. F. Mitrovi\'c, and C. Petrovic, \textit{J. Magn. Magn. Mater.} \textbf{307}, 301 (2006). 

\bibitem{Huy2009}  H. T. Huy, S. Noguchi, N. Van Hieu, X. Shao, T. Sugimoto, and T. Ishida, \textit{J. Magn. Magn. Mater.} \textbf{321}, 2425 (2009).

\bibitem{vanHieu2007}  N. Van Hieu, T. Takeuchi, H. Shishido, C. Tonohiro, T. Yamada, H. Nakashima, K. Sugiyama, 
R. Settai, T. D. Matsuda, Y. Haga, M. Hagiwara, K. Kindo, S. Araki, Y. Nozue, and Y. \=Onuki, \textit{J. Phys. Soc. Jpn.} \textbf{76}, 064702 (2007).

\bibitem{Schmiedeshoff2006}  G. M. Schmiedeshoff, A. W. Lounsbury, D. J. Luna, S. J. Tracy, A. J. Schramm, S. W. Tozer, V. F. Correa, S. T. Hannahs, T. P. Murphy, E. C. Palm, A. H. Lacerda, S. L. Bud'ko, P. C. Canfield, J. L. Smith, J. C. Lashley, and J. C. Cooley, \textit{Rev. Sci. Instrum.} \textbf{77}, 123907 (2006). 

\bibitem{Callen1965} E. Callen and H. B. Callen,  Physical Review \textbf{139}, A455 (1965).

\end{thebibliography}
\end{document}